\documentclass[reprint,superscriptaddress,amssymb,aps, pra, twocolumn,graphicx]{revtex4-1}

\usepackage{graphicx}
\usepackage{dcolumn}
\usepackage{bm}
\usepackage{hyperref}

\usepackage{amsmath} 
\usepackage{mathrsfs}
\usepackage[cmintegrals]{newtxmath}
\usepackage[mathlines]{lineno}

\usepackage{graphicx}
\usepackage{dcolumn}
\usepackage{bm}

\begin{document}

\preprint{APS/123-QED}

\title{Optomechanical compensatory cooling mechanism with exceptional points}

\author{Guo-Qing Qin}
\email{These authors contributed equally to this work}
\affiliation{Beijing Institute of Radio Measurement, The second Academy of China Aerospace Science and Industry Corporation (CASIC), Beijing 100854, China}
\author{Xuan Mao}
\email{These authors contributed equally to this work}
\affiliation{Department of Physics, State Key Laboratory of Low-Dimensional Quantum Physics, Tsinghua University, Beijing 100084, China}

\author{Hao Zhang}
\affiliation{Purple Mountain Laboratories, Nanjing 211111, China}

\author{Peng-Yu Wen}
\affiliation{Department of Physics, State Key Laboratory of Low-Dimensional Quantum Physics, Tsinghua University, Beijing 100084, China}

\author{Gui-Qin Li}
\affiliation{Department of Physics, State Key Laboratory of Low-Dimensional Quantum Physics, Tsinghua University, Beijing 100084, China}
\affiliation{Frontier Science Center for Quantum Information, Beijing 100084, China}
\author{Dong Ruan}
\affiliation{Department of Physics, State Key Laboratory of Low-Dimensional Quantum Physics, Tsinghua University, Beijing 100084, China}
\affiliation{Frontier Science Center for Quantum Information, Beijing 100084, China}
\author{Gui-Lu Long}
\email{gllong@tsinghua.edu.cn}
\affiliation{Department of Physics, State Key Laboratory of Low-Dimensional Quantum Physics, Tsinghua University, Beijing 100084, China}
\affiliation{Frontier Science Center for Quantum Information, Beijing 100084, China}
\affiliation{Beijing Academy of Quantum Information Sciences, Beijing 100193, China}

\date{\today}

\begin{abstract}
The ground state cooling of Brillouin scattering optomechanical system is limited by defects in practical sample. In this paper, we propose a new compensatory cooling mechanism for Brillouin scattering optomechanical system with exceptional points (EPs). By using the EPs both in optical and mechanical modes, the limited cooling process is compensated effectively. The dual-EPs system, which is discovered in this work for the first time, can be induced by two defects with specific relative angles and has function of not only actively manipulating the coupling strength of optical modes but also the Brillouin phonon modes.  Our results provide new tools to manipulate the optomechanical interaction in multi-mode systems and open the possibility of quantum state transfer and quantum interface protocols based on phonon cooling in quantum applications.
\end{abstract}

\maketitle


\section{introduction}
The coherent control of interaction between light and matter is at the heart of quantum technology and fundamental study. Cavity optomechanics \cite{aspelmeyer2014cavity} taking advantages of ultrahigh quality (Q) factor and small mode volume witnesses remarkable progress \cite{forsch2020microwave, ockeloen2018stabilized, riedinger2018remote, bienfait2019phonon, dong2012optomechanical, liu2013dynamic,  wilson2015measurement, xu2020nanoscatterer} over the last few decades. Based on microcavities, these platforms show great potential in different applications such as ground state cooling \cite{H2019Nonreciprocal, wilson2007theory, marquardt2007quantum, wilson2007theory, lai2018simultaneous, gan2019intracavity,  liu2022ground, Ke2021Quantum}, entanglement \cite{vitali2007optomechanical, liao2016macroscopic, zhong2020entanglement,   flayac2014heralded}, quantum state transfer \cite{stannigel2010optomechanical,  wang2012using, tian2012adiabatic}, nonreciprocity \cite{tang2022quantum, xu2018optomechanically, zhang2021nonreciprocal, 2020Optical}. Among quantum information processing tasks, high efficiency ground state cooling promises great initialization and fidelity \cite{stannigel2010optomechanical, stannigel2012optomechanical, riedinger2016non}. Several schemes have been proposed to realize tremendous performance of ground state cooling with various technologies \cite{arcizet2006radiation, park2009resolved, bahl2012observation}. With the increased phonon lifetimes and high frequency, Brillouin cavity optomechanics open the possibility for macroscopic quantum control even in room temperature. Based on coherent light-sound coupling, Stimulated Brillouin scattering has enabled high-efficiency laser cooling \cite{bahl2012observation, kim2017dynamically, otterstrom2018optomechanical, chen2016brillouin}. Furthermore, Brillouin system with a large optomechanical coupling rate provides a rich avenue for quantum mechanical state control at the single-phonon and multi-phonon level, which offers optomechanical practical interfaces in hybrid quantum network.

However, optomechanical cooling performance with whispering gallery mode (WGM)optical microcavities is limited by different factors in the real applications. Usually, Q factors of both optical and mechanical modes are important and the efforts of precise engineering to achieve ultrahigh Q factors have been made. Moreover, the defects such as deformed cavity geometry and scatters \cite{kim2019dynamic, nockel1994q, yu2015mechanism, griffith2015silicon} induced in the fabrication processes is also a major and easily overlooked constraint. These defects even in the subwavelength disorder can lead effective internal mode coupling \cite{mazzei2007controlled} with undesirable modes, which can not be eliminated for the state-of-the-art approaches. These internal modes scattering forms dark mode, which boundaries the ultimate limits of phonon cooling. The backscattering between clockwise (CW) and counterclockwise (CCW) optical modes can hinder the application performance in radiation-pressure cavity optomechanics. While in the Brillouin scattering optomechanical system, the situation is even worse where the internal defects scattering occurs between the two degenerate Brillouin mechanical modes. 

In this paper, we propose a new compensatory cooling mechanism for Brillouin scattering optomechanical system with exceptional points (EPs)\cite{chen2017exceptional, wiersig2014enhancing, de2022non, xu2016topological, qin2021experimental, jiang2022exceptional, ding2022non, zhang2018phonon, lu2018optomechanically, mao2020enhanced}. By using the EPs both in optical and mechanical modes, the limited cooling process is compensated effectively and we can overcome the cooling rate and limit in virtual optomechanical devices with fabrication imperfections. The dual-EPs system, which is discovered in this work for the first time, can be induced by two defects with specific relative angles and has function of not only actively manipulating the coupling strength of optical modes but also the Brillouin phonon modes. When both optical modes and mechanical modes are at EPs, the dark modes effect is eliminated. Our results provide new tools to manipulate the optomechanical interaction in multi-mode systems\cite{wiersig2014enhancing, xu2016topological} and open the possibility of quantum state transfer and quantum interface protocols based on phonon cooling in quantum applications.

The paper is organized as follows. In section \ref{section2} we describe the Brillouin scattering optomechanical system. In section \ref{section3} we analyze the optomechanical compensatory cooling. The conclusion is given in section \ref{section4}.

\begin{figure}
    \centering
    \includegraphics[width=\linewidth]{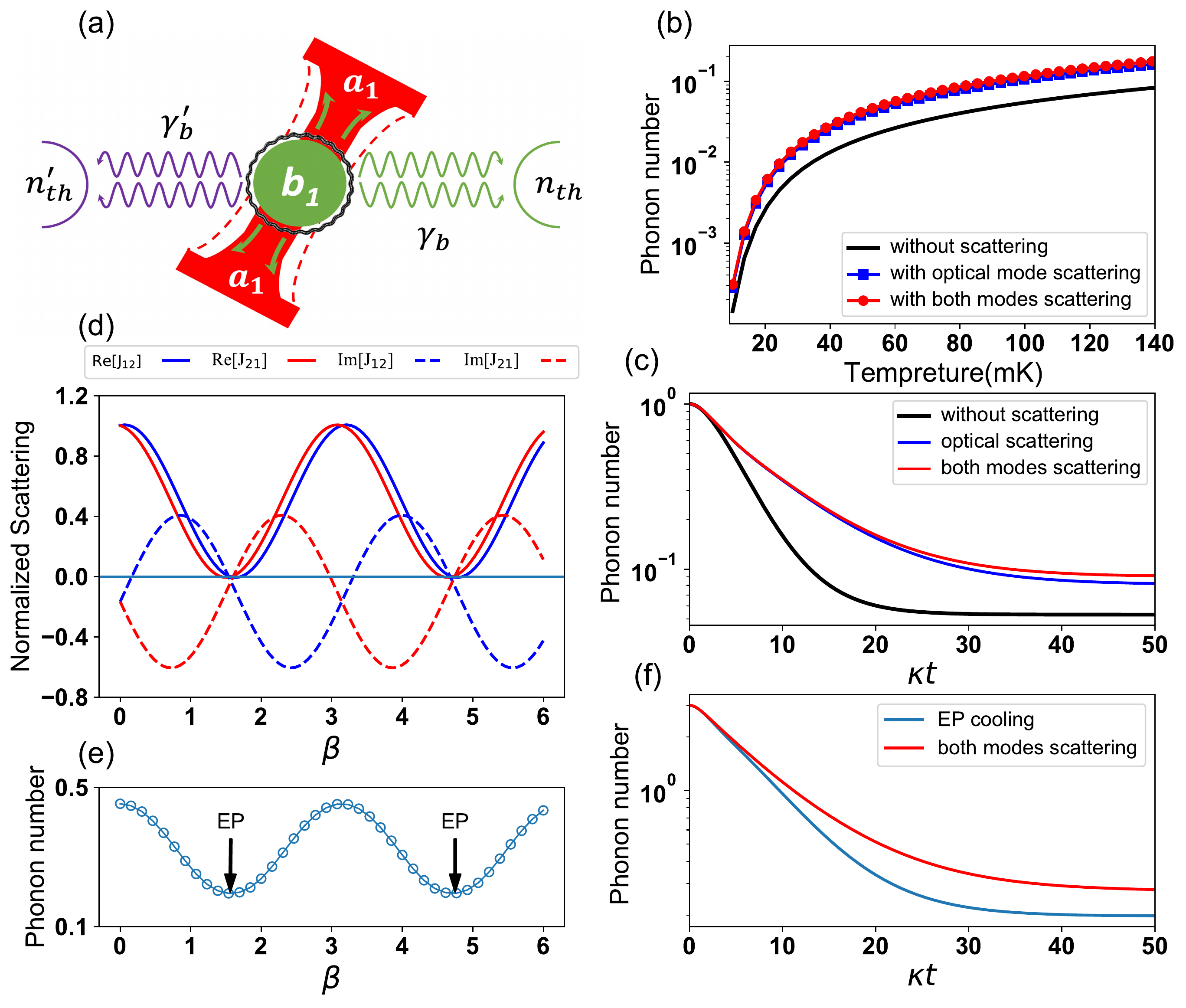}
    \caption{(a) Schematic of the optomechanical system with decrescent optomechanical coupling rate and an external effective thermal bath. (b) The steady-state phonon number for different mode scattering strength with $\omega_m = 183$MHz. (c) Dynamical evolution of the  phonon number with internal modes coupling when the thermal temperature is 79.5 mK. (d) The elements $J_{12}$, $J_{21}$ of the effective optical coupling and (e) the steady-state phonon number versus the the phase $\beta$ with $V=(0.35-0.1i)\gamma_o$, $\widetilde{n}_{th} =3$, $J_1 = J_2 = (0.502 -0.063i)\kappa_o $. (f) The time evolution of the phonon number when both optical and mechanical modes of the undriven system are prepared at EPs. The other parameters are $\kappa_o$= 1MHz, $J = (0.5-0.1i)\kappa_o$, $\gamma_o = 0.01\kappa_o$, $G = 0.24\kappa_o $.}
    \label{figure1}
\end{figure}  

\section{Brillouin scattering optomechanical system \label{section2}}
For a Brillouin scattering process, the Hamiltonian describing the optomechanical system with internal mode scattering is 
\begin{align}
    H =& \sum_{j} \Delta a_j^\dagger a_j + \sum_{j}\omega_{m} b_j^\dagger b_j + G (a_1^\dagger b_1 + b_1^\dagger a_1) \nonumber\\
    &+ J (a_1^\dagger a_2 + a_2^\dagger a_1)+ V (b_1^\dagger b_2 + b_2^\dagger b_1), \label{equation 1}
\end{align}
where ($\hbar = 1$), $a_j$($b_j$) denotes the optical (mechanical )mode with damping rate $\kappa_o$ ($\gamma_o$), $j= 1, 2$ label the CW and CCW mode, respectively. $\Delta$ is the detuning between the optical frequency and the pump laser, $G$ denotes the effective optomechanical strength. According to the Hamiltonian in Eq. \ref{equation 1}, the cooling performance of the mechanical mode $b_1$ in the weak coupling regime is limited by the internal modes scattering even for small $J$ and $V$ 
\begin{align}
    n_e &=\frac{1+C_{bb}}{1 + \frac{C_{ab}}{1+C_{aa}}+C_{bb}} \widetilde{n}_{th}.
    \label{equation 2}
\end{align}
Here, $\Delta = -\omega_{m}$, the cooperativity $C_{ab} = 4|G|^2 / \gamma \kappa$, $C_{aa} = 4|J|^2 / \kappa^2$, $C_{bb} = 4 |V|^2 / \gamma^2$, where $\kappa$ ($\gamma$) is the total optical (mechanical) loss with the scattering \cite{lu2018optomechanically}. In the calculation process, we neglect the photon occupation number in the environment since the optical resonant frequency is  orders of magnitude larger. When $J=0$ and $V=0$, the system is in the ideal condition with the classical cooling limit $\frac{\widetilde{n}_{th}}{C_{ab}}$. While in the real system with defects, the internal coupling would form dark states. These dark states hinder the cooling through different mechanisms seen in Eq. \ref{equation 2}. For example, when there is only internal coupling between optical modes ($V=0$), the interaction Hamiltonian can be rewritten in the form $\tilde{G} (a_1^\dagger b_{B} + b_{B}^\dagger a_1)$, in which the bright mode $b_{B} = (G b_1+ J a_2)/\tilde{G}$ is the  hybridization of mode $b_1$ and $a_2$ with coupling rate $\tilde{G} =\sqrt{|G|^2+|J|^2} $. Since the dark mode  $b_{d} = (J b_1- G a_2)/\tilde{G}$ decouples with mode $a_1$, the cooling efficiency is suppressed. In this case, the steady-state phonon number $n_e$ can be simplified under the condition of small $J$, $n_e = \frac{\widetilde{n}_{th},}{1 + C_{ab}(1-2C_{aa})}$. It is clear that the mechanism of deteriorative cooling origins from the decrescent optomechanical coupling rate induced by the optical internal coupling. The steady phonon number increases due to the dark state compared with ideal optomechanical device, shown in Fig. \ref{figure1} (b). This cooling suppression phenomenon imposes more restrictions on the cooling techniques and the thermal bath temperature. To understand the the evolution of the system with optical dark mode, the master equation is applied. Fig. \ref{figure1}(c) plots the dynamical cooling process with ideal (black line), only optical scattering (blue line). It is clear the suppressed optomechanical damping rate reduces the cooling speed and the system need more time to achieve the steady state. While in the case that the system suffers the coupling of optical modes $J=0$, dark state consisted of two mechanical modes emerges. We can calculate the phonon occupation and the fundamental limit can be simplified as $n_e =\frac{1+C_{bb}}{C_{ab}} \widetilde{n}_{th}$. The dark state with two mechanical modes leads to another heating channel for the system which poses a fundamental limit for laser cooling in virtual environment. The four modes coupled system described by the Eq. \ref{equation 1} can be treated as two modes coupled system with two effective mechanical thermal baths, shown in Fig. \ref{figure1} (a). The steady and dynamical phonon number of the system with both modes scattering is plotted in Fig. \ref{figure1} (b) and (c), respectively.

To achieve ground state cooling, suppressing thermal heating and enhance the optomechanical damping rate should be employed simultaneously. Different with the straightforward previous techniques which aims to eliminate the scattering, we will employ active control to the scattering to increase the optomechanical damping rate and suppress the heating swap between two mechanical modes. Especially, when the undriven system is prepared on the EPs, the dark mode effects can be eliminated. In this case, the interaction Hamiltonian of the undriven system is 
\begin{align}
    H_u = J_{12} c_1^\dagger c_2 + J_{21}  c_2^\dagger c_1 + V_{12}  b_1^\dagger b_2 + V_{21} b_2^\dagger b_1. \label{equation 5}
\end{align}
For the optical modes,  $J_{12} = J+J_1e^{2i\beta}$ ($J_{21}=J+J_2e^{-2i\beta}$) is the tunable coupling rate after actively controlling of the scattering with $J_1$($J_2$).  Fig. \ref{figure1} (d) demonstrates the tunability of the optical mode coupling strength and all the elements of $J_{12}$ and $J_{21}$ vanish for certain phase $\beta$. The best cooling performance is achieved when the undriven system is at optical EPs, shown in Fig. \ref{figure1} (e). At the same time, the $V_{12}$ ($V_{21}$ ) has the same form of $J_{12}$ ($J_{21}$) and the energy scattering of mechanical modes could be tuned to achieve mechanical EPs through $V_1$ and $V_2$.  When both optical modes and mechanical modes of the undriven system are prepared at EPs by setting $V_{1} = V_{2} = (0.35-0.068i)\gamma_o$, the cooling  efficiency is further optimized, shown in Fig. \ref{figure1}(f).

\begin{figure}
    \centering
    \includegraphics[width=\linewidth]{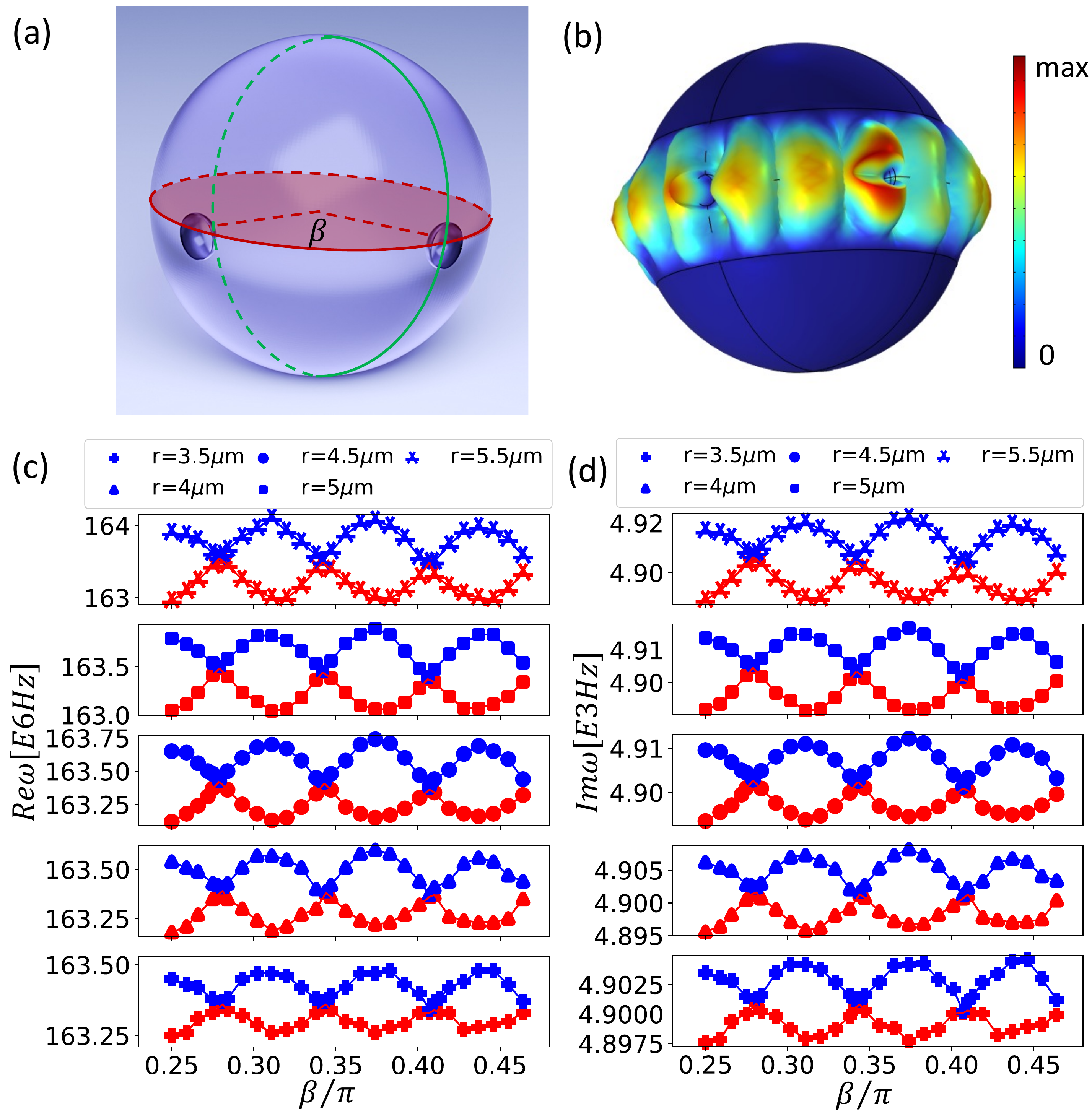}
    \caption{(a) Diagram of micro-resonator with two defects, the relative angle between the two defects is $\beta$. (b) Simulation result of the typical mechanical vibration characterized by eigen-frequency around $163$ MHz. (c), (d) The real parts and the imaginary parts of the mechanical eigenvalues with different defect sizes versus the relative angle $\beta$. The radius of the resonator $R = 60 \mu m$.}
    \label{mechanical}
\end{figure}

\begin{figure}
    \centering
    \includegraphics[width=\linewidth]{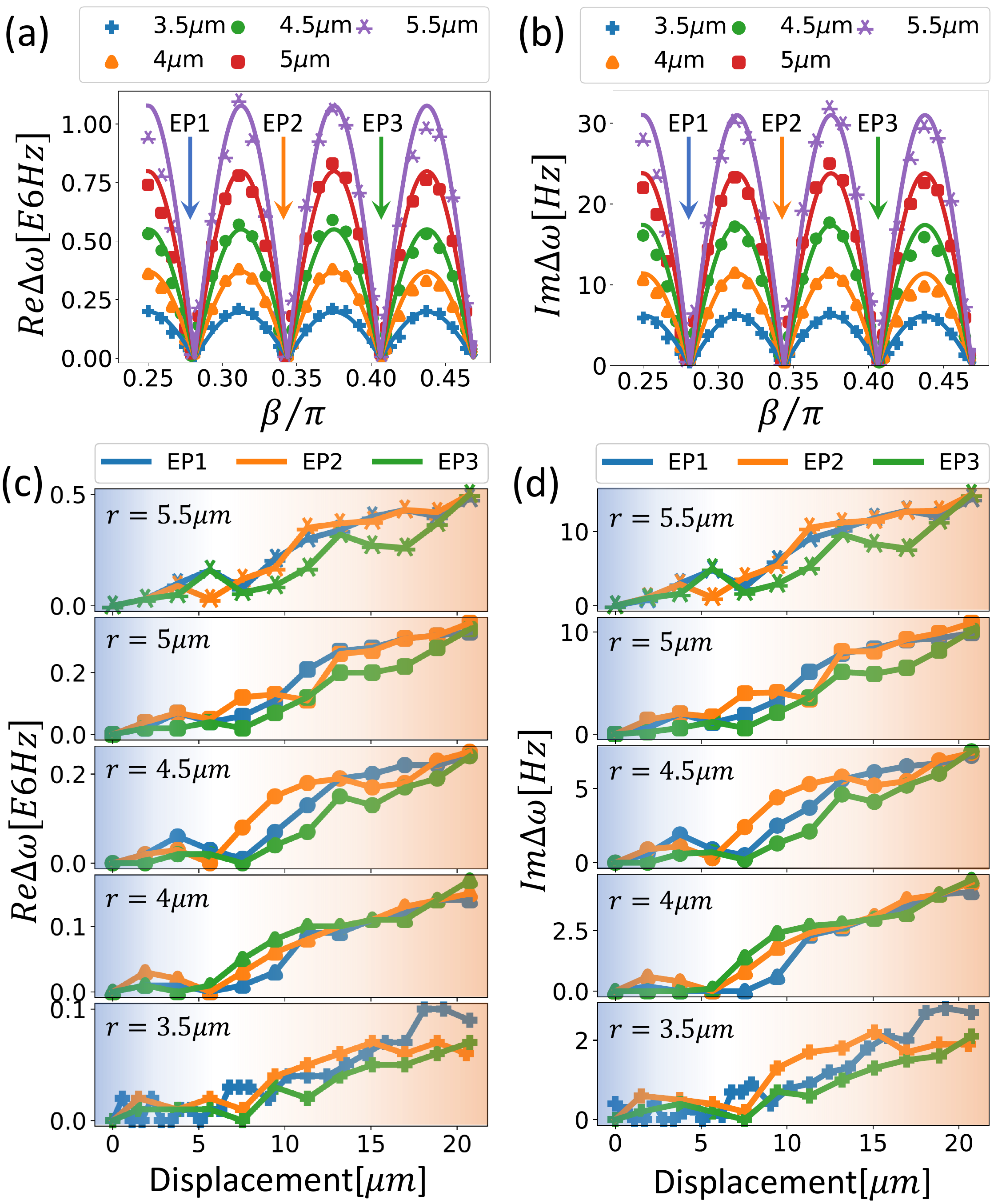}
    \caption{(a), (b) The real parts difference and the imaginary parts difference of the mechanical eigenvalues versus the relative angle for different defect sizes. (c), (d) The real parts difference and the imaginary parts difference of the mechanical eigenvalues versus the displacement along the green line in Fig. \ref{mechanical} at three EPs for different defect sizes. The radius of the resonator $R = 60 \mu m$. The mode coupling rates of the defect size $3.5 \mu m$, $4 \mu m$, $4.5 \mu m$, $5 \mu m$, and $5.5 \mu m$ are corresponding to $\{e_{b1}, e_{b2}\} = \{5 \times 10^4 - 1.5 i, 4.9 \times 10^4 - 1.6 i\}$Hz, $\{9.3 \times 10^4 - 2.8 i, 9.2 \times 10^4 - 2.9 i\}$Hz, $\{13.8 \times 10^4 - 4.3 i, 13.7 \times 10^4 - 4.4 i\}$Hz, $\{20 \times 10^4 - 5.9 i, 19.9 \times 10^4 - 6 i\}$Hz, and $\{27 \times 10^4 - 7.7 i, 26.9 \times 10^4 - 7.8 i\}$Hz, respectively.}
    \label{robust}
\end{figure} 

\section{Optomechanical compensatory cooling \label{section3}}
In order to realize a more effective cooling device containing the main features of the toy model where the coherent control of modes scattering is applied, we take the advantage of defects to form EPs. The universal design is shown in Fig. \ref{mechanical} (a). The two defects with the relative angle $\beta$ are introduced to form EPs in the resonator which consist of optical mode pair ($a_1$, $a_2$) with degenerate frequency $\omega_a$ and mechanical mode pair ($b_1$, $b_2$) with degenerate frequency $\omega_m$. We implement finite-element simulations of the defected resonator with different defect sizes. The typical simulation result of the Brillouin mode and the mechanical vibration distribute around the equatorial plane of the resonator with azimuthal number $m_b =8$ is shown in Fig. \ref{mechanical} (b). Note that this set of parameters serves as an example and there is no principle limitation in resonator size, defect size, wavelength, or refractive indices. Fig. \ref{mechanical} (c) and \ref{mechanical} (d) demonstrate the real parts and the imaginary parts of the mechanical eigenvalues, which corresponding to the frequencies and the linewidths, with different defect sizes. It can be found that the frequencies and the linewidths of the Brillouin mode pair share the similar behavior which exhibits periodicity as $\beta$ changes and the period is $\pi/m_b$. 

Fig. \ref{robust} (a) and \ref{robust} (b) denote the real parts difference and the imaginary parts difference with different defect sizes. Also, the differences $Re\Delta \omega$ and $Im\Delta \omega$ change periodically with $\beta$ and have the same period $\pi/m_b$, $\Delta\omega = 2\sqrt{e^2_{b1}+e^2_{b2}+2 e_{b1} e_{b2} cos(2m_b\beta)}$. $e_{b1}$ and $e_{b2}$ are the mode coupling rates induced by the defect 1 and defect 2, respectively. Furthermore, the larger the defects is, the larger the maximum value of the difference reaches. The simulation results have a good agreement with the qualitative analysis. Both the coupling coefficients $J_{b1}$ and $J_{b2}$ have exponential factor $2 i m_b \beta$ which determines the period is $\pi/m_b$. The influence of the defects on the Brillouin mode pair is restricted by their size. Small defects show a limit impact on the mode which causes slight difference of the real parts and imaginary parts. At some specific points, i. e. $0.279 \pi$, $0.343 \pi$, and $0.407 \pi$, the real parts and the imaginary parts of the Brillouin mode pair are coalescent simultaneously which are called EPs. To test the robustness of the defected system, the position of the defect 2 is applied a displacement from the equatorial plane along the green line shown in Fig. \ref{mechanical} (a). For different defect sizes, the real parts difference and the imaginary parts difference at three EPs angle versus the displacement are demonstrated in Fig. \ref{robust} (c) and \ref{robust} (d). It can be found that at some displacement range ($\leq 5 \mu m$) the EPs are maintained and the system show robustness with the position displacement along the direction perpendicular to equatorial plane. Furthermore, the robustness of the system is better for smaller defect size.

\begin{figure}
    \centering
    \includegraphics[width=\linewidth]{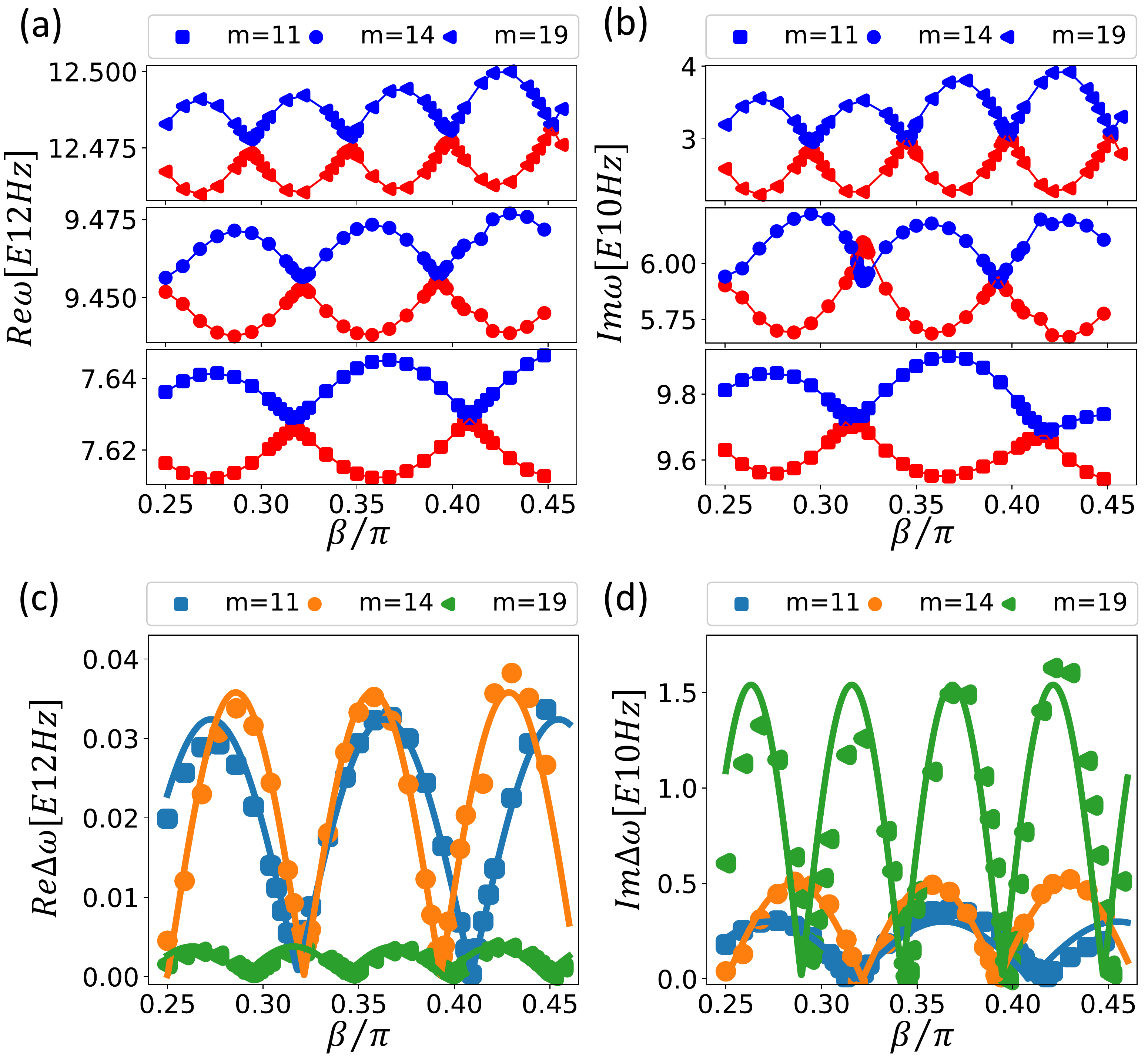}
    \caption{(a), (b) The real parts and the imaginary parts of the optical eigenvalues with different azimuthal mode numbers versus the relative angle $\beta$. (c), (d) The real parts difference and the imaginary parts difference of the optical eigenvalues versus the relative angle $\beta$ with different azimuthal mode numbers. The dots in this figure is the simulation result and the solid lines in (c) and (d) are the theoretical curves. The radius of the cavity is $R = 60 \mu m$ and the radius of the defects is $r = 4.5 \mu m$. The mode coupling rates of $m = 11$, $14$, and $19$ are corresponding to $\{e_{a1}, e_{a2}\} = \{8.3 \times 10^9 - 7 \times 10^8 i, 7.9 \times 10^9 - 8 \times 10^8 i\}$Hz, $\{9 \times 10^9 - 1.2 \times 10^9 i, 8.9 \times 10^9 - 1.3 \times 10^9 i\}$Hz, and $\{1 \times 10^9 - 3.8 \times 10^9 i, 0.9 \times 10^9 - 3.9 \times 10^9 i\}$Hz, respectively.}
    \label{optical}
\end{figure}

The proposed scheme also works to reach optical EP and the visualized mode control upon defects engineering is shown in Fig. \ref{optical}. The azimuthal number of the optical modes consider here are $m_a = 11$, $14$, and $19$. Fig. \ref{optical} (a) and \ref{optical} (b) demonstrate the real parts and the imaginary parts of the optical eigenvalues with different azimuthal numbers as $\beta$ varies. The frequencies and the linewidths of the optical modes have the same period $\pi/m_c$ which is predicted by the theoretical analysis. Also, there are EPs for the optical mode pair at which the frequencies and the linewidths are coalescent simultaneously. For example, $\beta = 0.452 \pi$ is an EP for $m_a = 19$ case. To demonstrate the presence and the absence of EPs intuitively, Fig. \ref{optical} (c) and \ref{optical} (d) imply the difference between the real parts and the imaginary parts of the optical modes. 

 \begin{figure}
    \centering
    \includegraphics[width=\linewidth]{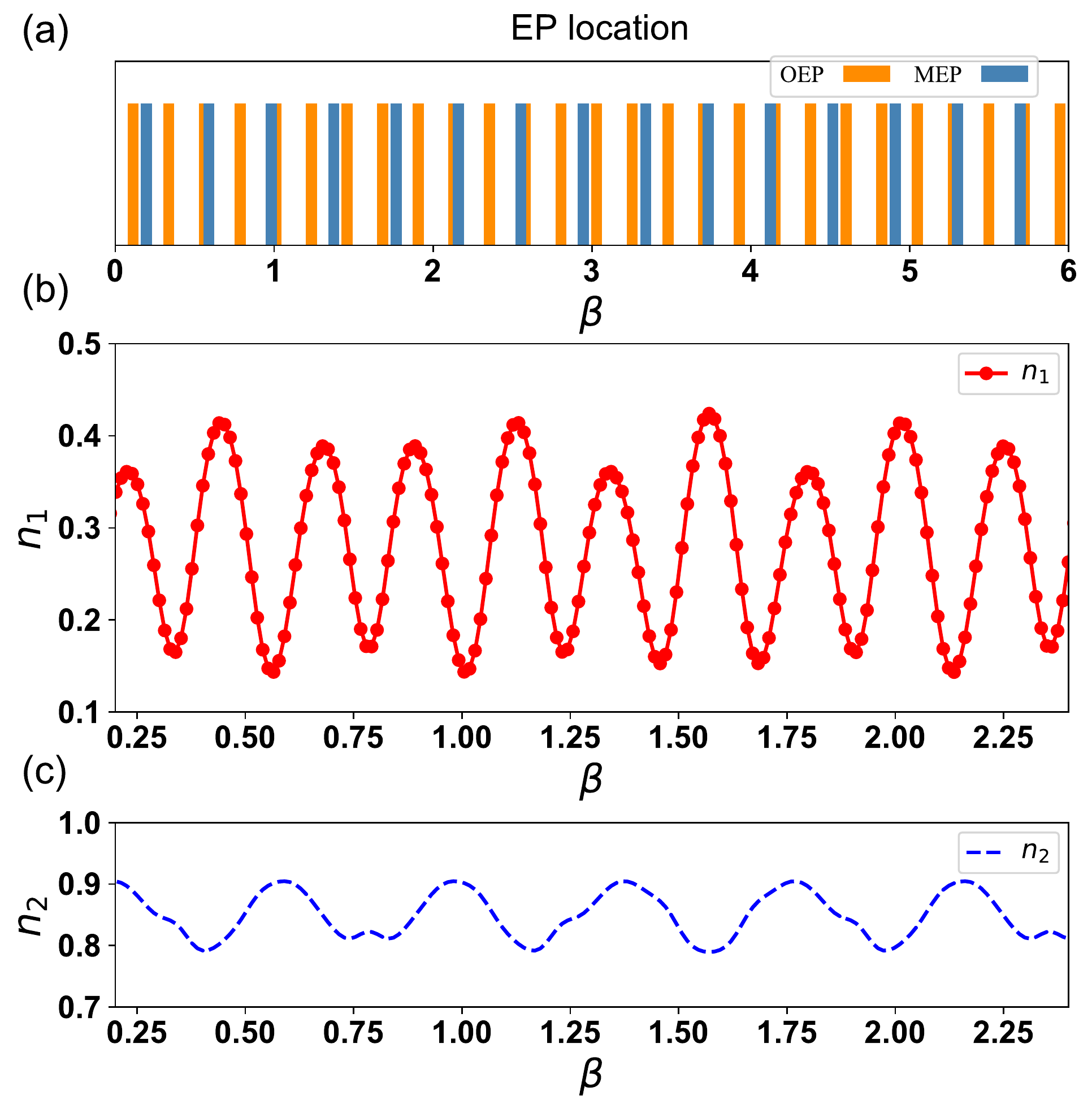}
    \caption{(a) The normalized steady phonon number of two mechanical modes versus phase $\beta$. (a) The OEPs and MEPs emerge for different defects location. (b) and (c) plot the normalized occupancy of mode $b_1$ and $b_2$ with optomechanical coupling strength $G/\kappa_o  = 0.155$, $m_a = 14$, $m_b = 8$. The other parameters are the same as used in Fig. \ref{figure1}. }
    \label{figure5}
\end{figure} 

Now we study the cooling  performance of the mechanical resonator using the optomechanical system describe above. To calculate the evolution for all modes of the optomechanical system, we should take all the dissipations induced by the two defects into consideration. To demonstrate the impact of the defects on cooling, we calculate the steady solution of two mechanical modes for various angle of $\beta$, shown in Fig. \ref{figure5}. Due to the azimuthal numbers of the optical and mechanical modes are not the same, the periods of mechanical exceptional point (MEP) and optical exceptional point (OEP) are different in Fig. \ref{figure5}(a). This MEP and OEP occur at different angle $\beta$ can affect the phonon occupancy. The plots show that both mode $b_1$, $b_2$ represented by red dotted line and blue dashed line are cooled when the relative angle changes. For mechanical mode $b_1$, the steady phonon number also demonstrates periodic variation. In general, the period is $\frac{m \pi}{m_a m_b}$ where $m$ is the least common multiple of $m_a$ and $m_b$ and in this case is $\frac{\pi}{2}$. When $\beta = 0.336$, the phonon number also decreases to its local minimum. However, the mechanical modes still couples with each other resulting in suppressing the cooling of $b_1$. While $\beta = 0.577$, the undriven optical modes and mechanical modes are prepared at their EP and the phonon number of cooled mode $b_1$ reaches the global minimum. The phonon number of $b_2$ is reaching the global maximum, indicating that the cooling of $b_2$ is efficiently suppressed. The dark mode effect is completely suppressed and the phonon number reaches its cooling limit.

\section{conclusion \label{section4}}

We have proposed a new compensatory cooling mechanism for Brillouin scattering optomechanical system with EPs. By using the EPs both in optical and mechanical modes, the limited cooling process is compensated effectively as EPs can suppress thermal heating and enhance the optomechanical damping rate simultaneously. The dual-EPs system, which is discovered in this work for the first time, can be induced by two defects with specific relative angles and has function of not only actively manipulating the coupling strength of optical modes but also the Brillouin phonon modes. We study the efficient design of the optomechanical device which present great feasibility and robustness. This approach can overcome the ultimate limits of phonon cooling induced by mode scattering in real and fundamental applications. Furthermore, our design can be used to realize nonreciprocity and quantum state engineering in macroscopic systems.

\section*{acknowledgements}
This work is supported by the National Natural Science Foundation of China (61727801, 62131002), National Key Research and Development Program of China (2017YFA0303700), the Key Research and Development Program of Guangdong province (2018B030325002), Beijing Advanced Innovation Center for Future Chip (ICFC), and Tsinghua University Initiative Scientific Research Program.


\nocite{*}



%

\end{document}